\documentclass{aa}
\usepackage{txfonts}
\usepackage{graphicx}
\usepackage{natbib}
\usepackage[colorlinks]{hyperref}
\hypersetup{
  colorlinks=true,
  linkcolor=blue,
  filecolor=magenta,      
  urlcolor=cyan
}
\begin{document}  

\title{Collision-induced satellite in  the  blue wing of the
  Balmer-$\beta$ line and consequences on the Balmer series}

\author{
  F.~ Spiegelman\inst{1}
  \and  
  N.~F.~Allard   \inst{2,3}
  \and
  J. F. Kielkopf  \inst{4}
}
   \institute{Laboratoire de Physique et Chimie Quantique, F\'ed\'eration FERMI,
 Universit\'e de Toulouse (UPS) and CNRS, 118 route de Narbonne, 
           F-31400 Toulouse,   France \\
\and
   GEPI, Observatoire de Paris,  Universit\'e PSL, 
     UMR 8111, CNRS,
     61, Avenue de l’Observatoire, F-75014 Paris, France\\
          \email{nicole.allard@obspm.fr}\\
     \and
Sorbonne Universit\'e, CNRS, UMR7095, Institut d'Astrophysique
de Paris, 98bis Boulevard Arago, PARIS, France\\
          \and
Department of Physics and Astronomy, 
    University of Louisville, Louisville, Kentucky 40292 USA \\
}

\date{Received 22 november 2021 / Accepted 3 january 2022} 

\abstract{In this paper we  emphasize the non-Lorentzian behavior of the Balmer series in helium-dominated DBA white dwarf stars for which the   decades-old problem exists  for the determination of the hydrogen abundance. In a very recent work, we have shown that  quasi-molecular line satellites due to H-He and H-H collisions are responsible for the asymmetrical shape of the Lyman-$\alpha$ lines observed with the {\it Cosmic Origin Spectrograph} ({\it COS}) and that a similar asymmetry  exists for the Balmer-$\alpha$ line profiles. In continuation with  very recent work, where the $n=2, 3$ potential energies and transition dipole moments from the ground state were determined, here, we present  accurate H-He potential energies and electronic transition dipole moments concerning the molecular states correlated with H($n$=4)+He and their transition dipole moments with the states correlated with H($n$=2)+He.  Those new data are used  to provide a  theoretical investigation  of the collisional effects in the blue wing of the  Balmer-$\beta$ line of H perturbed by He. Because of the general trend  characterizing the repulsive $\Sigma$ states of the potential energies involved in  the Balmer series,   the amplitude in the core of the line is decreasing very fast with the order of the series when the helium density gets as large as 10$^{21}$ cm$^{-3}$. This study is undertaken by applying  a unified theory of spectral line broadening that is valid at very high helium densities found in DZA white dwarf stars.
  The treatment includes collision-induced (CI) line satellites due to asymptotically forbidden transitions, and it explains the asymmetry observed in their spectra.
}


\titlerunning{Collisional-induced satellite in  the  blue wing of
  Balmer-$\beta$ line}
\authorrunning{Spiegelman, Allard \&  Kielkopf}

\maketitle

\section{Introduction}
\label{sec:introduction}

In the spectra of white dwarf stars exhibiting strong heliums lines with weaker hydrogen lines, that is to say DBA white dwarfs,  there is a discrepancy between  the  values of the hydrogen abundances
determined from  Balmer-$\alpha$  in the visible and those
from  Lyman-$\alpha$ in the ultraviolet
\citep[see][and references therein]{xu2017}.
The existence of  close
line satellites in the blue wing of these lines was found in
detailed 
collisional broadening profiles computed for both
H-He and H-H \citep{spiegelman2021,allard2021}. 
These features are responsible for the asymmetrical shape and the
decreasing  strength of the core of the lines in those cases.
In this paper we extend our study to the  Balmer-$\beta$ line by considering the H($n$=4) perturbed by He. 

In recent previous works, we used
new  multi-reference configuration interaction (MRCI) calculations of the  
 excited state potential energy curves  dissociating into H($n$=1,2,3)+He, 
 as well as  the relevant electric dipole transition moments
 contributing  to the Lyman-$\alpha$ and Balmer-$\alpha$ spectra.
 We  illustrated how  
 tiny relativistic effects affect the 
 asymptotic correlation of the H-He adiabatic states and change the related
 transition dipole moments.
 
In the present  work, we determine potential energy curves dissociating
 into H($n$=4)+He  
 and transition dipole moments of H-He  involving initial and final states
 correlated with H($n$=2)+He and H($n$=4)+He, respectively.  
Between these states, there are  23   H-He transitions which generate
the complete Balmer-$\beta$ line profile.
The   $\Sigma-\Sigma$ transitions  provide the essential contribution
to the blue wing of the Balmer series lines, explain their asymmetrical behavior, and are relevant to the  cool DBA white dwarf analyses. The potential energy curves and transition dipole moments are discussed in Sect.~\ref{sec:potHHe}.
Unified profile calculations are presented in Sect.~\ref{sec:BaHHe}.
They show that in He-rich white dwarf stars, the non-Lorentzian shape
of Balmer-$\beta$ increases with He density and it will affect abundance
analyses if not included in stellar atmosphere models. 
 Consequently the strength and shape of the Balmer line series are also diagnostics of 
the atmospheric conditions of cool DZA white dwarf stars.

In the Sect.~\ref{sec:trend}  we consider 
 the general trend of the repulsive $\Sigma$ state of the
excited states of H-He. They contribute through the Lyman and Balmer lines and
lead to a non-Lorentzian shape that is increasingly significant in higher series members that affect the visible spectrum  of helium-rich white dwarfs.

\section{Diatomic H-He potentials and electronic transition dipole moments}
\label{sec:potHHe}

\begin{figure}
\centering
\vspace{8mm}
\includegraphics[width=8cm]{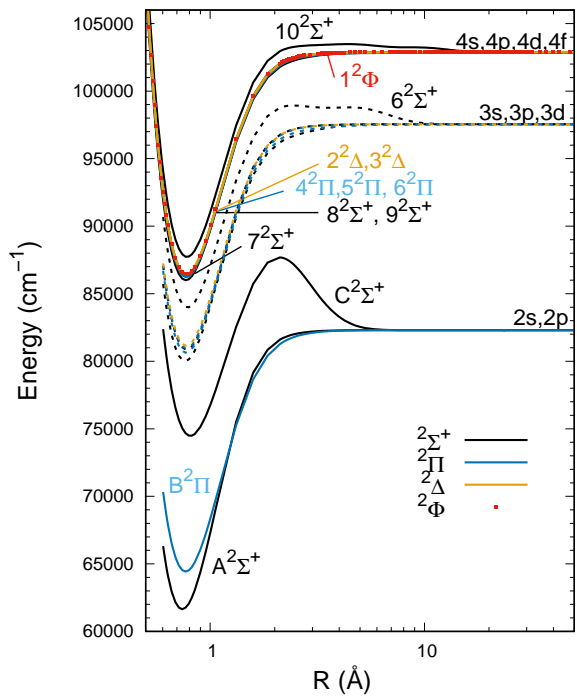}
\caption{MRCI adiabatic potential energy curves of molecular states
  dissociating into  H($n$=2,4)+He. The $R$ scale is a logarithmic graduation.
  The dashed lines correspond to potential energy curves dissociating into  
  H($n$=3)+He.}
\label{fig:pothhe}
\end{figure}
The calculation scheme for the potentials dissociating
into H($n$=2,3,4)+He is the same as described in 
our previous publications   concerned with the  Lyman-$\alpha$ 
\citep{spiegelman2021} and the Balmer-$\alpha$ \cite{allard2021}
broadenings. Briefly, a 
MRCI \citep{knowles92,molpro2015} calculation was run within an extensive Gaussian-type
orbital (GTO) basis set.
Scalar relativistic corrections (namely the Darwin and mass-velocity terms) according to the Douglas-Kroll-Hess (DKH) scheme \citep{reiher2006,nakajima2011}, breaking the specific
degeneracy of the atomic hydrogen levels determined with the Coulomb
Hamiltonian only, were added. 
In order to improve the description of the  adiabatic molecular states of H-He toward
the $n=4$ atomic limit, 
we refined the basis set used in our previous
works \citep{spiegelman2021,allard2021},
consisting of an aug-cc-pv6z  quality description complemented by diffuse exponents for all $s$, $p$, $d$, $f$, $g$,  and $h$ functions for H and He,
reoptimizing the exponents of the external $d$ and $f$ GTO functions of hydrogen
(see Table~A1 of the appendix).  Table~A2 of the
appendix  compares the  calculated transitions  
from $1s$ to $n$=4
with  the  $j$-averaged experimental ones. 
The calculated $4s-4f$ splitting is 0.14~cm$^{-1~}$ larger, but  of the same order of magnitude as the experimental value 0.061 cm$^{-1}$.

\begin{figure}
\centering
\vspace{8mm}
\includegraphics[width=8cm]{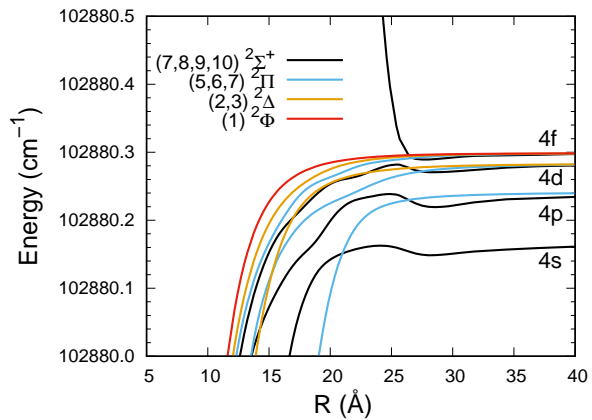}
        \caption{Long range zoom on the adiabatic potential energy curves of H($n$=4)+He.}
\label{fig:e4lr}
\end{figure}

In the following, we  use the  spectroscopist's notations $X$, $A$, $B$, and $C$
to label the  lower adiabatic molecular states of H-He,  namely the ground state $1^2\Sigma^+$, the lowest excited states $2~^2\Sigma^+$,
$1~^2\Pi$, and $3~^2\Sigma^+$ correlated with H($n$=2)+He, while we
label  the upper states correlated with H($n$=4)+He according to their 
adiabatic ranking in their  respective symmetries.
For a detailed characterization of states $n$=2,3, we refer readers to our previous papers \citep{spiegelman2021,allard2021} and 
to earlier literature \citep{theodora1987,
ketterle85,ketterle88,ketterle89,sarpal1991,lo2006,vanhemert91,ketterle90a,ketterle90b,allard2020}.
The molecular states dissociating into $4s$, $4p$, $4d$, and $4$f are
(7-10)~$^2\Sigma^+$, (4-6)~$^2\Pi$, (2,3)~$^2\Delta$, and $1~^2\Phi$. 
Figure~\ref{fig:pothhe} shows the potential energy curves of the states
dissociating into H($n$=2)+He and H($n$=4)+He.
All states are bound with equilibrium distances close to
0.77~\AA\/,  most of them have similar dissociation energies around
16500~cm$^{-1}$ and vibrational constants $\omega_{exe}$, and they can hardly be
individually identified at the scale of the figure.  
Those dissociation energies are also quite similar to those of the HeH$^+$
ion ($R_e$=0.7742~\AA\/, $D_e$=16455.64~cm$^{-1}$, and $\omega_e$=3220~cm$^{-1}$)
calculated 
by \citet{kolos76}, which are consistent with the
Rydberg nature of those states \citep{ketterle90c,ketterle90d}.
The detailed spectroscopic constants 
are given in the Table A~3 of the appendix.  The most stable state  is $7~^2\Sigma^+$ with a dissociation energy of $D_e$=16882~cm$^{-1}$. Conversely, state 10~$^2\Sigma^+$ is  the least stable ($D_e$=15157~cm$^{-1}$), and it has a barrier to dissociation.
This feature is analog to those  observed for the respective highest states
dissociating  into the
 $n=2$ and $n=3$ configurations, namely $C ^2\Sigma^+$ and  $6 ~^2\Sigma^+$.
To our knowledge, no previous theoretical data  exist  for the molecular states of HHe correlated with $n=4$. A detailed experimental  investigation was published by \citet{ketterle90c} who could characterize states $4~^2\Pi$, $5~^2\Pi$, and $2~^2\Delta$.
The present calculated  $v'=0$ to $v"=0$ vibrational transitions
 ($T_{00}$ values) are in agreement with Ketterle's data with a $rms$
deviation of 16~cm$^{-1}$ (see appendix, Table~A.3). At long distance in
the range 17-28~\AA\/, the states  undergo multiple avoided
crossings, which determine their correlation  with the atomic asymptotes,
as illustrated in Fig.~\ref{fig:e4lr}.

\begin{figure}
\centering
\vspace{8mm}
\includegraphics[width=8cm]{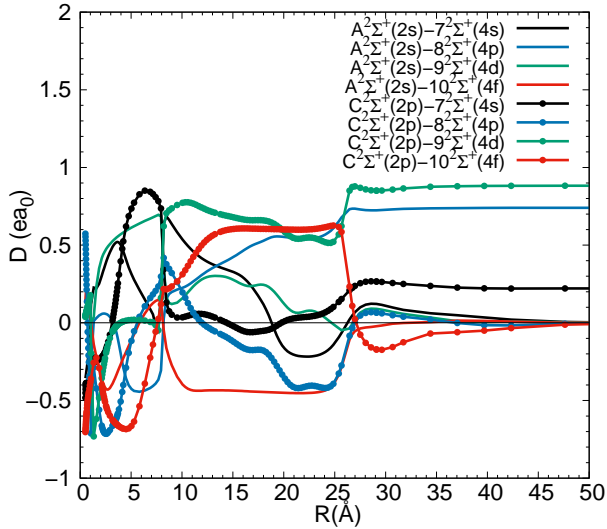}
        \caption{Transition dipole moments $D(R)$ of H-He between  
         $^2\Sigma^+$ states $n$=2 and  $n$=4.}
\label{fig:dipsig24}
\end{figure}

All transition dipole moments $D(R)$ between the $n$=2 and $n$=4 adiabatic
states were calculated using the MRCI wavefunctions, namely transitions
from $A~^2\Sigma^+$, $C~^2\Sigma^+$ to 
7,8,9,10~$^2\Sigma^+$ and to 4,5,6~$^2\Pi$,  from  $B~^2\Pi$ to
7,8,9,10 ~$^2\Sigma^+$, 4,5,6~$^2\Pi$ and  2,3~$^2\Delta$.
The \mbox{$A^2\Sigma^+, C ^2\Sigma^+, B ^2\Pi- 10 ^2\Sigma^+$}
 asymptotically forbidden
transitions are involved in the blue wing contribution, as discussed below.
The $A^2\Sigma^+, C ^2\Sigma^+ - 7,8,9,10 ^2\Sigma^+$ transition moments
  display a rather complex picture
shown in Fig.~\ref{fig:dipsig24}. Their evolution can be rationalized considering the following three key features: (i) the avoided crossing between 
 the lower states $A~^2\Sigma^+$ 
and $C~^2\Sigma^+$ around R=8.1~\AA\/ \citep{spiegelman2021}, (ii) the multiple avoided crossing between the upper states $(7-10)^2\Sigma^+$ 
in the range  $R$=17-28~\AA\/ illustrated in Fig. \ref{fig:e4lr}, and (iii) other avoided crossings between the upper states at short distance of $R \le 5$~\AA. 

The situation is thus  similar to what was observed for the Lyman-$\alpha$ and Balmer-$\alpha$ molecular transitions:
while relativistic effects, partly taken into account here, obviously
play a negligible role in the energies of the  H($4f$)+He states
(less than $\approx$ 0.1 cm$^-1$)  
and in the transition dipole moments at  short distance, they break the Coulomb asymptotic degeneracy  and  induce
cascade avoided crossings which determine the asymptotic correlation of the adiabatic states and the
variation of the dipole transition moments at  medium and long distance.

\begin{figure}
\centering
\vspace{8mm}
\includegraphics[width=8cm]{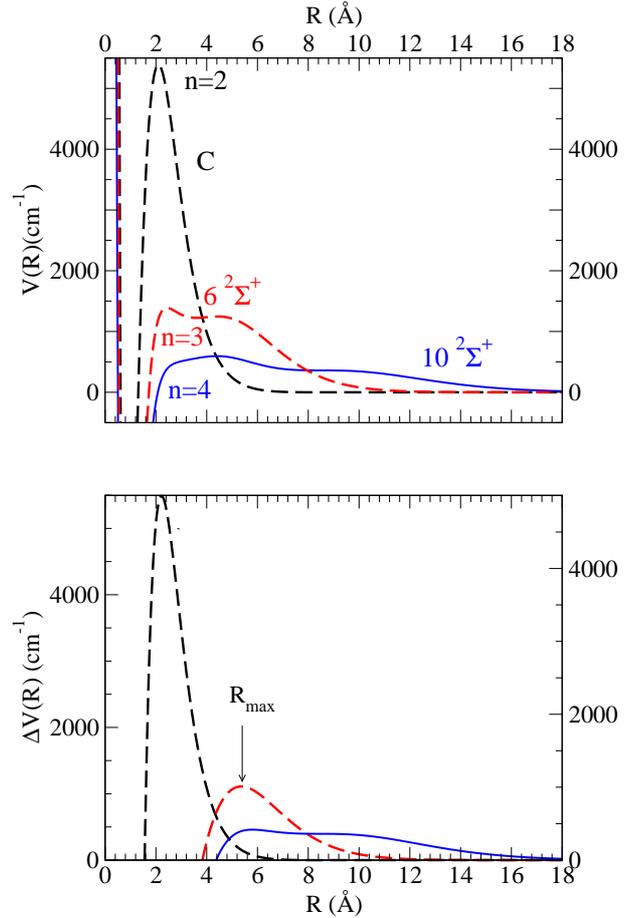}
\caption{ Repulsive potentials and their difference.
  Top: Potential energy $V$ of the repulsive
  states of  Lyman-$\alpha$ (dashed black curve),
  Balmer-$\alpha$ (dashed red curve),
  and Balmer-$\beta$ (blue curve).
  Bottom: Difference  potential $\Delta V (R)$ for the Lyman-$\alpha$
  transition
  \mbox{$X \rightarrow C$} (dashed black curve),
 for the Balmer-$\alpha$ transition
\mbox{$C \rightarrow 6^2\Sigma^+$} (dashed red curve),
and for the Balmer-$\beta$ transitions 
\mbox{$C \rightarrow 10^2\Sigma^+$} (blue curve).}
\label{fig:Erepuls}
\end{figure}

\section{Analysis of the  Balmer-$\beta$ profile perturbed by  He atoms}
\label{sec:BaHHe}

The highest molecular states correlated to $n$=2 and $n$=3 have a prominent
role in the appearance of blue satellite features in  Lyman-$\alpha$
\citep{spiegelman2021} and Balmer-$\alpha$ \citep{allard2021} line
profiles.
If we define the potential energy $V(R)$ for a state $e$  as
\begin{equation}
V_{e}(R) = E_e(R)-E_e^{\infty} \; ,
\label{eq:V}
\end{equation}
the prediction  of a line satellite in the blue wing 
 is related to the  maximum of  $V(R)$ of these  repulsive states.
The upper box of Fig.~\ref{fig:Erepuls} shows a comparison of  the short-range part of
the  potential curve  $V(R)$ of the following  repulsive states:
\mbox{$C~^2\Sigma^+$, 6~$^2\Sigma^+$, and 10~$^2\Sigma^{+}$}, 
 corresponding to the highest states correlated   to the
$n=2$, $n=3$, and $n=4$ levels, respectively.
With increasing $n$, the 
barrier height decreases (5386~cm$^{-1}$, 1390~cm$^{-1}$, and 574~cm$^{-1}$
 for $n$=2, 3, and 4, respectively) and the repulsion of the state above
 the asymptote rises at greater distances, namely $R$=8, 15, and 25~\AA\/ ~for
 $n$=2, 3, and 4,  as can be seen in
 Figs.~\ref{fig:pothhe}-\ref{fig:e4lr}.

It is essential to use  a general unified theory in which the electric dipole moment varies
during a collision because all transitions of Balmer-$\beta$ from the 2$s$ and 2$p$ states 
to the $4f$ state are  asymptotically forbidden.
 Most of the problems in collision-induced radiative transitions have been
solved within the one-perturber approximation. At low densities, the 
binary model for an optically active atom in collision with one perturber
is valid for the whole profile, except for the central part of the line.
 In dense plasmas, as in very cool DZ  white dwarfs
  \citep {allard2016a,kawka2021} and liquid
  helium clusters \citep {allard2013a},
the possibility of several atoms interacting strongly is high, and the effects
play a role in the wavelength of the line center, for example in the shift of the line as
well as the general shape of the line profile. 
For such a high  perturber density, the  collisional effects
should be treated by using the autocorrelation formalism in order
to take simultaneous collisions with more than one perturbing
atom into account.
 A pairwise additive assumption
 allows us to calculate the total profile $I(\Delta \omega)$, when
all the perturbers interact through the Fourier transform (FT)  of the $N^{\rm th}$ power of the
autocorrelation function $\phi (s)$ of a unique atom-perturber
pair. Therefore
\begin{equation}
\Phi(s)=(\phi(s))^{N}\; .
\end{equation}
That is to say, we neglect the interperturber correlations. The radiator
can interact with several perturbers simultaneously, but the perturbers do
not interact with each other. It is what~\citet{royer1980} 
calls the "totally uncorrelated perturbers approximation''.
The fundamental result expressing the autocorrelation
function for many perturbers in terms
of a single perturber quantity $g(s)$ was first obtained
 by~\citet{anderson1952} and~\citet{baranger1958a} 
in the classical and quantum cases, 
respectively.
From the point of view of a general classical theory, the solution to
the ~\citet{anderson1952} model corresponds 
to the first order approximation
in the gas density obtained by the cumulant expansion method~\citep{royer1972}.
The higher order terms representing correlations 
between the perturbers  are neglected since they  are
 extremely complicated~\citep{royer1972,kubo1962,kubo1963,vankampen1974}.
In \citet{allard1999}, 
we  derived a classical path expression for a pressure-broadened atomic
 spectral line shape that allows for an electric dipole moment that is
 dependent on the position of perturbers, which is not included in the
 approximations of~\citet{anderson1952}
 and~\citet{baranger1958a,baranger1958b}.

 The spectrum $I(\Delta\omega)$ can be written as the 
FT of the dipole autocorrelation function $\Phi(s)$ ,
\begin{equation}
I(\Delta\omega)=
\frac{1}{\pi} \, Re \, \int^{+\infty}_0\Phi(s)e^{-i \Delta\omega s} ds,
\label{eq:int}
\end{equation}
where $s$ is time.
The FT in Eq.~(\ref{eq:int}) 
is taken such that $I$($\Delta\omega$) is
normalized to unity when integrated over all frequencies,
and $\Delta$ $\omega$ is measured relative to  the 
unperturbed line.
We have provided an overview
of  the unified theory in Sect.~3 of \citet{spiegelman2021}.

The  unified theory  predicts that line satellites
will be centered periodically at frequencies corresponding to integer
multiples of the extrema of  $\Delta V(R)$, the difference  between the
energies of the quasi-molecular transition. The $\Delta V(R)$ is defined as follows:

\begin{equation}
\Delta V(R) \equiv V_{e' e}(R) = V_{e' }(R) - V_{ e}(R) \; .
\label{eq:deltaV}
\end{equation}

\begin{figure}
\centering
\vspace{8mm}
\includegraphics[width=6cm,angle=270]{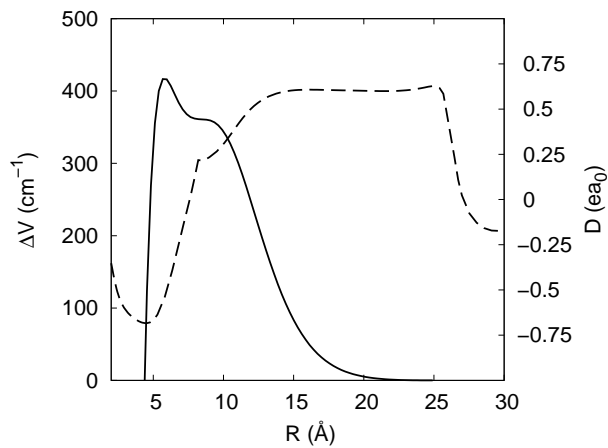}
\caption{Difference potential energy $\Delta V(R)$ in cm$^{-1}$  (black line)
and  dipole moment $D(R)$ (black dashed line) for  the forbidden
\mbox {$C~2p$ $\rightarrow$ 4$f$~10$^2\Sigma^+$} transition.}
\label{fig:POTDIP}
\end{figure}

\begin{figure}
\centering
\vspace{8mm}
\includegraphics[width=8cm]{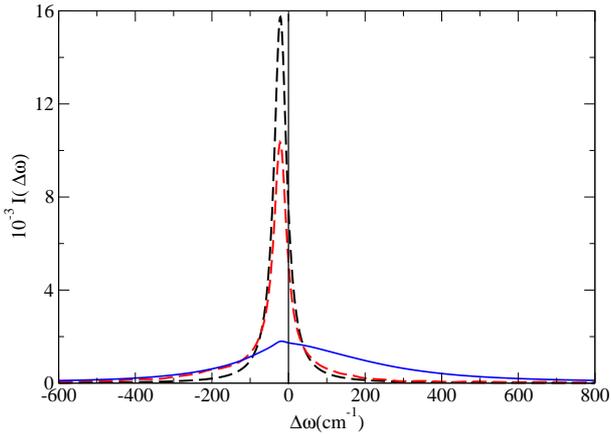}
\caption{Comparison  of the  total  profile (blue line) to the profile
  restrictly obtained with the allowed transitions (red dashed line).
  The Balmer-$\beta$ line profile is also compared to the
  Lorentzian approximation (black dashed curve).
  \mbox {The  He density is 5$\times$ 10$^{20}$~cm$^{-3}$},
and   the  temperature is 8000~K.}
\label{fig:BAHHeomg}
\end{figure}

\begin{figure}
\centering
\vspace{8mm}
\includegraphics[width=8cm]{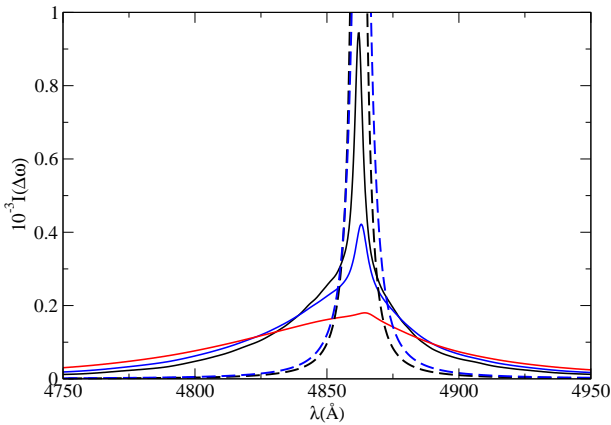}
\caption{Variation with the He density of the Balmer-$\beta$ line profile for
  2~$\times$ 10$^{20}$ (black line),
  3~$\times$ 10$^{20}$ (blue line),
  and  5~$\times$ 10$^{20}$~cm$^{-3}$ (red line);
  the  temperature is 8000~K.
 The Balmer-$\beta$ line profiles are also compared to the
  Lorentzian approximation for the  He densities
  2$\times$ 10$^{20}$~cm$^{-3}$ (dashed black line) and
  3$\times$ 10$^{20}$~cm$^{-3}$ (dashed blue line).}
\label{fig:BAHHelam}
\end{figure}

\begin{figure}
\centering
\vspace{8mm}
\includegraphics[width=8cm]{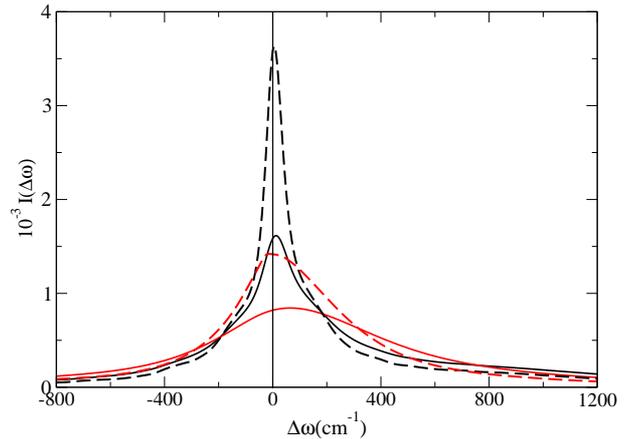}
\caption{Comparison  of the Balmer-$\alpha$ (black curves)
  and  Balmer-$\beta$ (red curves) line profiles.
  The  He densities are  10$^{21}$  cm$^{-3}$ (full curves) and
  6$\times$ 10$^{20}$  cm$^{-3}$ (dashed curves); the  temperature is 8000~K.}
\label{fig:alfabeta}
\end{figure}

The bottom  of Fig.~\ref{fig:Erepuls} shows $\Delta V(R)$ for the  transitions
\mbox{$X \rightarrow C$} related to Lyman-$\alpha$, 
\mbox{$C \rightarrow 6~^2\Sigma^+$} related to 
Balmer-$\alpha$ compared to $\Delta V(R)$ for the  transition
\mbox{$C \rightarrow 10~^2\Sigma^+$} related to Balmer-$\beta$.
The main contribution to the blue wing is due to the forbidden 
\mbox{$C\,2p~^2\Sigma^+$ $\rightarrow 4f~10^2\Sigma^+$} transition.
The transition dipole moment $D(R)$ between the $ C \, 2p $
electronic state and the  4$f$ state asymptotically vanishes as
the radiative transition between the corresponding  atomic states is forbidden
(Fig.~\ref{fig:dipsig24}).
 Although this transition should not contribute to the unperturbed line
profile,
radiative transitions can be induced by close collisions because $D(R)$  differs from zero when a He atom  passes close to the H atom.
To point out the importance of the variation of the dipole moment 
on the formation of   collision-induced (CI) satellite, 
we have displayed 
$D(R)$ together with the corresponding $\Delta V(R)$ for the  $ C \, 2p -4f$
 transition  in Fig.~\ref{fig:POTDIP}.
The satellite amplitude depends on the value of $D(R)$ through the  region
of the potential maximum responsible
 for the satellite and on the position of this extremum \citep{allard1998b}.
The presence of quasi-molecular satellites due to collision  effects
is unimportant when the He density is low;  however, in the physical conditions
found in the cool atmospheres of He-rich white dwarfs, their presence leads to
an asymmetrical shape of the Balmer profiles which  is a signature
of a high helium density.
 Figure~\ref{fig:BAHHeomg} shows
a very broad blue wing with a  close CI line  satellite
at about 200~cm$^{-1}$ from the line center corresponding to the
maximum of $\Delta V (R)$.
The line profile obtained when taking only the allowed transitions
into account is overplotted. The main objective  of  this figure is to illustrate the 
contribution of the forbidden transitions to the Balmer-$\beta$ profile.
Figure~\ref{fig:BAHHelam} shows that the
presence of a line satellite in the near line wing leads to a complex
behavior of the dependence of the line shape on He density.
 As in the case of of  the DZA white dwarf L745-46A obtained
  at the ESO La Silla 3.6m telescope, the observed spectrum of Ross 640
  obtained at Calar Alto (Fig.~7 in \citet{koester2000})
  shows an observed  Balmer-$\beta$ line with a very broad blue wing  due to
    high He densities which can reach 10$^{21}$~cm$^{-3}$
    in the Balmer-$\alpha$ and $\beta$ formation region
    (Vennes 2021, private communication). Fig.~\ref{fig:alfabeta}  shows a very 
    broad line profile at this He density for both 
    Balmer lines. In particular,
  we have found that the Balmer-$\beta$ line profile
  is totally asymmetric in this range of He density.
   This effect is already present at the lower density,
    2$\times$ 10$^{20}$~cm$^{-3}$, and it has an
     increasing importance with
     He density (Fig.~\ref{fig:BAHHelam}).
The development of the blue wing 
 leads to the center of the main line being overwhelmed by the blend of 
line satellites. As a result,
the full treatment of the Balmer-$\beta$  line 
reveals strong effects on its opacity outside the impact approximation.
 The Lorentzian approximation is overplotted in  Figs.~\ref{fig:BAHHeomg}-\ref{fig:BAHHelam}.
Consequently, we conclude that these collision-induced effects are not always
negligible when models of stellar spectra are compared with observations
to determine abundances, structural properties, or ages.

\section{Consequences on the Balmer series}
\label{sec:trend}
 
Figure~\ref{fig:pothhe} illustrates the potential energies involved in
Lyman-$\alpha$, Balmer-$\alpha$, and Balmer-$\beta$.
Due to the more diffuse
character of the 4$s$, 4$p$, 4$d$, and 4$f$ orbitals
with respect to those with smaller  $n$, the barrier height
of the highest state extends to a wider  distance range
and is lower than that of states $C~^2\Sigma^+$ and 6~$^2\Sigma^+$.
The characteristics of the highest states must be stressed
in view of the study of the asymmetrical shape dependence on
the order of the Balmer series.
The top of Fig.~\ref{fig:Erepuls}  summarizes the potential energies  of the
highest states involved in Lyman-$\alpha$, Balmer-$\alpha$, and
Balmer-$\beta$, and the bottom of it shows the corresponding $\Delta V(R)$.  The essential
characteristics  to note are that 
the maximum in $\Delta V$  occurs at larger
internuclear distances ($R_{\mathrm max}$ $\sim$ 10~\AA\/)
for Balmer-$\beta$ than for Balmer-$\alpha$ ($R_{\mathrm max}$ $\sim$ 5.4~\AA\/)
and Lyman-$\alpha$ ($R_{\mathrm max}$ $\sim$ 2~\AA\/). The
$\Delta V(R)$ presents a  flat maximum at a larger distance for Balmer-$\beta$.
The average number of perturbers in the
interaction volume at $R_{\mathrm max}$ is the most  determining parameter for
the amplitude of the satellites in the spectral line
\citep{allard1978,royer1978,allard1982}.
This dependence on the average number of
perturbers in the collision volume is expected on the basis
of the Poisson distribution, which indicates the probability
of finding a given  number of uncorrelated perturbers in
the collision volume.
It was  decisively identified  in the theoretical analysis of experimental
Cs spectra by \citet{kielkopf1979}.
The other crucial point is that  the maximum $\Delta V_{\mathrm max}$ is
smaller with  350~cm$^{-1}$ for Balmer-$\beta$,
versus 1000~cm$^{-1}$  for the Balmer-$\alpha$ line and 
 5000 cm$^{-1}$ for the Lyman-$\alpha$ line. 
 The Balmer-$\beta$ line satellite gets closer  to the main line than
 the Balmer-$\alpha$ and  Lyman satellites, as shown in 
 Fig.~\ref{fig:alfabeta}, Fig.~6 of \citet{allard2021}, and
 Fig.~11 of \citet{spiegelman2021}.
 Its effect on the asymmetrical shape of the Balmer profile and  on
 its amplitude becomes more important and appears at a lower density.

 Because of this general trend  characterizing the repulsive $\Sigma$ states,
 the average number of perturbers in the interaction 
 volume will get larger for higher series of Balmer lines,
  leading to a higher probability of multiple-pertuber
  effects.  The core of the main line due to allowed transitions
  will disappear, being   replaced by the blend of
  multiple satellites~\citep{kielkopf1983,kielkopf1985}.
  This is what is already happening in the Balmer-$\beta$ line when the
  He density gets as large as  10$^{21}$~cm$^{-3}$.
  Fig.~\ref{fig:alfabeta}  shows a  broad line profile
  with a decreasing asymmetry and a change in the sign of its shift
  resulting from the disappearance of the main line due to the allowed
  transitions replaced by the blend of multiple line satellites.
  Increasing the He density has the same effect as considering a
  higher member of the Balmer series as it increases the
  probability of multiple pertuber effects. 
  This behavior is well known in line broadening theory and has been
  thoroughly studied in the 1950s by the Ch'en group \citep{allard1982}.
  The  spectra of members of the Balmer series corresponding
    to a transition to upper level $n$ are then  progressively more blueshifted
    and shallower than the larger $n$.
    The observed spectra of PG 1157+004
    (Fig.~17 11598+007 of \citet{limoges2015}) or
  of WDJ0103-0522 (Fig.~2 of \citet{tremblay2020}, which are both 
  massive  white dwarfs with a very high log $g$, 
  exhibit a strong asymmetry and a blue shift, 
providing further evidence that we are dealing with  an unresolved blend of quasi-molecular
line satellites due to  high He densities.
 For PG 1157+004, the helium density  is  3x10$^{20}$ cm$^{-3}$
  in the  region of the  formation of the Balmer-$\alpha$
  and $\beta$ lines (Vennes 2021 private communication).
  Figure~\ref{fig:BAHHelam} shows that even at this rather low density, the
  Balmer-$\beta$ line profile shows an asymmetrical shape.
 For cooler atmospheres, as in the case of WDJ0103-0522,  the He density is as high as
10$^{21}$ (Vennes 2021 private communication).
Many explanations have tried to solve that puzzle of such white dwarf spectra, 
such as  two sources contributing  to the Balmer features
\citep{limoges2015} or  magnetic fields \citep{tremblay2020}.

The resonance broadening of H perturbed by collisions with H atoms
produces asymmetry in the Balmer-$\alpha$ line profiles similar to  that
due to H-He \citep{allard2021},  but the H density in DA white dwarfs
  is not as large as the 
He density in  cool He-rich DBA to have such an effect in their observed spectra.
This a major difference between  Balmer spectra of a DA or of  a cool DBA
  white dwarf.
More generally,  the important sensitivity of the Balmer line shapes
  to the nature of the perturber  and its density
suggests that they could be used as a  diagnostic tool  given precision low-noise observed spectra and appropriate line shape theory based on the high accuracy now provided by first-principles theoretical atomic and molecular physics.

In conclusion, an  accurate determination of neutral-collision-broadened lines is required to interpret 
the strength, broadening, and shift of the resulting Balmer line profiles correctly.  Calculations reported in \citet{allard2021} for Balmer-$\alpha$ and in this paper for Balmer-$\beta$, especially for H-H collisions, involve a huge number of  transitions, and they are dependent on very accurate molecular data.  The  past laboratory experimental work and analysis of observations of multiple-perturber effects on the broadening of atomic spectral lines such as by \citet{kielkopf1983,kielkopf1985} is also fundamental, as we have pointed out because it confirms  that the higher the excitation of the atomic state, the larger the effects of multiple perturbers are at a lower density in the laboratory \citep{kielkopf1979}, as  shown in Fig.~\ref{fig:alfabeta} for stellar atmospheres.

\begin{acknowledgements}
    We  would like to thank Adela Kawka and Stephane Vennes
    who initiated this study.
    We also acknowledge with appreciation Stephane Vennes,
    for his  determinations of He density in the Balmer-$\alpha$  and $\beta$
    formation regions of the DZ white dwarfs which are cited in the paper.
    We thank the anonymous referee  for helpful comments that improved the
    manuscript.
\end{acknowledgements}


\begin{appendix}
\onecolumn
        \section{Complementary H-He molecular data}
        \begin{center}
                {{\bf Table~A.1}. Complementary basis set information}\\

        \begin{tabular}{cl}
        \hline
                \cal{l} & \hspace*{1cm}     exponents\\
        \hline
                    &\\
                $d$ &4.453000, 1.958000, 0.861000, 0.378000, 0.139000, 0.058407,\\
                    &0.024364, 0.010714, 0.005000, 0.002550, 0.000850\\
                    &\\
                $f$ &4.100000, 1.780000, 0.773000, 0.345000, 0.126396, 0.049205,\\
                     &0.019065, 0.008000, 0.003700, 0.001600,\\
                    &\\
        \hline
                    &\\
        \end{tabular}\\
\vspace{-0.2cm}
        \footnotesize{Exponents of the optimized $d$ and $f$ Gaussian-type functions of Hydrogen}
        \label{tab:basis}
        \end{center}
\vspace{1cm}
\begin{center}
        {{\bf Table~A.2}. Calculated atomic energy levels of Hydrogen ($n=4)$ $vs$ experiment}        \\
\begin{tabular}{cccc}
\hline
        $level$  & Coulomb /DKH&Coulomb/DKH/mass & experiment\\
\hline
        &&&\\
        4s &102880.164&102824.160& 102823.853\\
        4p &102880.240 &102824.236& 102823.882\\
        4d &102880.283&102824.279 &102823.903\\
        4f &102880.299&102824.295 &102823.914 \\
\hline
        &&&\\
\end{tabular}\\
        \end{center}
\vspace{-0.5cm}
\footnotesize{Transition energies  of hydrogen  (in cm$^{-1})$ from $1s$ to  the
  $n$=4 levels: theoretical levels including the DKH
  contribution (second column), theoretical levels with DKH contribution
        and finite proton mass correction (third column), and $(2j+1)$-weighted averages of experimental terms \citep{nist2020}
        (fourth column).} 
\vspace*{1cm}
        \begin{center} 
                {{\bf Table~A.3}. Spectroscopic constants of HHe molecular states correlated with H($n$=4)+He}\\
        \begin{tabular}{ccccccccc}
          \hline
          State & ref & $R_e$ (\AA) & $\omega_e$ (cm$^{-1})$ & $\omega_{exe}$ (cm$^{-1}$) & $\omega_{eye}$ (cm$^{-1}$) & $D_e$ (eV/cm$^ {-1}$) & $T_e$ (cm$^{-1}$) & $T_{00}$ (cm$^{-1}$)\\
                \hline
        &&&&&&&\\
        $7^2\Sigma^+$& (a)& 0.770022  & 3304.0 & 156.4  & 49.3 & 2.093/16882.5 & 24352.2 & 24140.1  \\ 
        $8^2\Sigma^+$& (a)& 0.775062  & 3235.4 & 176.2  & 24.7 & 2.052/16556.7 & 24678.1 & 24423.7  \\ 
        $9^2\Sigma^+$& (a)& 0.774535  & 3242.1 & 176.9  & 23.9 & 2.043/16480.1 & 24754.8 & 24503.5  \\ 
        $10^2\Sigma^+$&(a)& 0.779162  & 3202.1 & 173.3  & 26.9 & 1.879/15157.6 & 26077.8 & 25807.8  \\ 
        $4^2\Pi$&      (a)& 0.773681  & 3253.3 & 177.8  & 23.1 & 2.065/16651.8 & 24582.9 & 24336.9  \\ 
                      &(b)&           &        &        &      &               &         & 24335.01  \\
        $5^2\Pi$&      (a)& 0.774614  & 3242.8 & 176.0  & 24.1 & 2.048/16517.9 & 24717.0 & 24466.2  \\
                      &(b)&           &        &        &      &               &         & 24441.18  \\
        $6^2\Pi$&      (a)& 0.774488  & 3242.7 & 177.0  & 23.9 & 2.043/16474.7 & 24760.7 & 24509.7  \\
        $2^2\Delta$&   (a)& 0.774384  & 3244.0 & 177.0  & 23.8 & 2.041/16461.3 & 24773.5 & 24523.1  \\
        &(b)&           &        &        &      &               &         & 24539.01  \\
        $3^2\Delta$&   (a)& 0.774116  & 3247.3 & 177.2  & 23.6 & 2.038/16436.5 & 24798.9 & 24550.1  \\
        $1^2\Phi$&     (a)& 0.774229  & 3245.9 & 177.1  & 23.6 & 2.038/16441.6 & 24793.8 & 24544.3  \\
        &&&&&&&\\
        \hline
        &&&&&&&\\
        \end{tabular}\\
        \end{center}
\vspace{-0.5cm}
        \footnotesize{Spectroscopic constants of HHe molecular states dissociating into  H(4s,4p,4d,4f)+He(1s$^2$).
         Electronic dissociation energies $D_e$ were  calculated  with reference to  the respective dissociation limits. We note that 
         $T_e$ is the adiabatic  electronic transition energy from state $A^2\Sigma^+$, and $T_{00}$ is the associated 
          $v'=0$ ~to ~$v"=0$ transition energy.   (a) theory, this work; (b) experiment \citep{ketterle90c}.}
        \label{tab:spectro}
\end{appendix}

\end{document}